\documentclass[reprint,secnumarabic,amssymb,amsmath,nobibnotes,aps,prd] {revtex4-1}
\usepackage[all, knot]{xy}

\begin{document}

\title{The modification of the energy spectrum of charged particles by exotic
open 4-smoothness via superstring theory}

\author{Torsten Asselmeyer-Maluga}

\email[REV\TeX{} Support: ]{torsten.asselmeyer-maluga@dlr.de}

\affiliation{German Aero space center, Rutherfordstr. 2, 12489 Berlin}

\author{Pawe\l{}\, Gusin}

\email[REV\TeX{} Support: ]{pawel.gusin@us.edu.pl}

\affiliation{University of Silesia, Institute of Physics, ul. Uniwesytecka 4,
40-007 Katowice}

\author{Jerzy Kr\'ol}

\email[REV\TeX{} Support: ]{ iriking@wp.pl}

\affiliation{University of Silesia, Institute of Physics, ul. Uniwesytecka 4,
40-007 Katowice}

\begin{abstract}
In this paper we present a model where the modified Landau-like levels
of charged particles in a magnetic field are determined due to the
modified smoothness of $\mathbb{R}^{4}$ as underlying structure of
the Minkowski spacetime. Then the standard smoothness of $\mathbb{R}^{4}$
is shifted to the exotic $\mathbb{R}_{k}^{4}$, $k=2p$, $p=1,2...$.
This is achieved by superstring theory using gravitational backreaction
induced from a strong, almost constant magnetic field on standard
$\mathbb{R}^{4}$. The exact string background containing flat $\mathbb{R}^{4}$
is replaced consistently by the curved geometry of $SU(2)_{k}\times\mathbb{R}$
as part of the modified exact backgrounds. This corresponds to the
change of smoothness on $\mathbb{R}^{4}$ from the standard $\mathbb{R}^{4}$
to some exotic $\mathbb{R}_{k}^{4}$. The calculations of the spectra
are using the CFT marginal deformations and Wess-Zumino-Witten (WZW)
models. The marginal deformations capture the effects of the magnetic
field as well as its gravitational backreactions. The spectra depend
on even level $k$ of WZW on $SU(2)$. At the same time the WZ term
as element of $H^{3}(SU(2),\mathbb{R})$ determines also the exotic
smooth $\mathbb{R}_{k}^{4}$. As the consequence we obtain a non-zero
mass-gap emerges in the spectrum induced from the presence of an exotic
$\mathbb{R}_{k}^{4}$. 
\end{abstract}

\date{August 30, 2010}

\maketitle
\tableofcontents{}

\section{Introduction}

For every charged quantum particle, say $e$, moving through a non-flat
gravitational background there should exist a high energy limit where
one has to include quantum gravitational effects. Additionally this
particle cannot be described by perturbative field theory any longer.
In dimension 4 quantum gravity (QG) is still far from being a complete
theory. The classical part is well understood by using (pseudo-)Riemannian
smooth geometry but the transformation of classical (geometric) gravity
to QG is a mystery especially from the purely mathematical point of
view. In this paper we address both issues and show how this transformation
can be done by a (special) smooth Riemannian 4-geometry.

QG is already defined in 10 spacetime dimensions as superstring theory
(though not as any complete quantum field theory). There exist many
techniques worked out in superstring theory which refer to dimension
4 like: compactification, flux stabilization, brane configuration
model-buildings, brane worlds, holography or AdS/CFT and some others.
But the transition from higher dimensional string theory to 4-dimensional
physics using these methods is rather ambiguous. Recently two of us
have proposed \cite{AssKrol2010ICM,AsselmeyerKrol2011,AsselmeyerKrol2011b,AsselmKrol2011d}
to incorporate 4-dimensional effects into string theory via its connections
with an inherently 4-dimensional differential geometric phenomenon:
exotic smoothness of topologically trivial $\mathbb{R}^{4}$. Among
all $\mathbb{R}^{n}$ only the case $n=4$ allows for different smoothings
of the Euclidean space, i.e. an atlas (or a system of reference frames)
non-diffeomorphic to the usual global coordinate patch of the $\mathbb{R}^{n}$
(one chart). Especially we proposed how to transfer some string theory
techniques for backgrounds containing an exotic $\mathbb{R}^{4}$
as 4-dimensional part. String theory for these backgrounds define
in a new way the connections with 4-dimensional physics. This new
4-dimensional window of superstring theory refers as a tool to exact
backgrounds in any order of $\alpha$'. Such backgrounds are rather
exceptional, though very desirable, in superstring theory (see e.g.
\cite{OrlandoCorfu06,Orlando2006}). We will make extensive use of
these backgrounds also in this paper and show direct applicability
of the link between \emph{strings} and \emph{4-exoticness} in deriving
physical results in dimension 4. Especially we will answer the questions:
Given exotic smooth differential structure on $\mathbb{R}^{4}$, i)
can one describe physical effects due to the propagation of strong
magnetic field on this exotic 4-geometry underlying Minkowski spacetime?
and ii) How can one deal consistently with gravity of non-flat exotic
metrics on $\mathbb{R}^{4}$ at the quantum level? The answers depend
strongly on the QG regime. In particular the QG backreaction of extremely
strong magnetic fields $H$ is evaluated. We obtain the energy spectra
of spin and spinless charged particles in a magnetic field on these
exotic backgrounds by superstring theory. The geometry used in these
calculations is induced by some exotic $\mathbb{R}^{4}$ carrying
both gravity regimes, classical gravity on a non-flat 4-manifold \cite{Sladkowski2001},
and the QG regime \cite{AssKrol2010ICM,AsselmeyerKrol2011}.

In superstring theory one can consistently grasp the QG effects by
changing the exact background containing the flat $\mathbb{R}^{4}$
to a background with a curved 4-dimensional part, i.e. $\mathbb{R}^{4}\times K^{6}\to SU(2)_{k}\times\mathbb{R}_{\phi}\times K^{6}$.
The $SU(2)_{k}\times\mathbb{R}_{\phi}$ part is the background of
the linear dilaton $\phi\sim QX^{0}$ in the superconformal theory
on $SU(2)\times\mathbb{R}$ \cite{CHS1991,KK95,AsselmeyerKrol2011}
where $Q$ is the charge. This change of the backgrounds allows us
to derive 4D effects from QG backreaction of strong (magnetic) fields.
These effects can be calculated by using methods from 2D CFT and $SU(2)_{k}$
WZW model \cite{KK95,SenHass93,Orlando2006}. Then the change of the
backgrounds corresponds to the change of smoothness on 4-space: $\mathbb{R}^{4}\to\mathbb{R}_{k}^{4}$
as global structure of the local Minkowskian spacetime. 

In our paper, we consider some energy spectra of particles on $S^{3}$
with the presence of a magnetic field (Subsec. \ref{sub:Flat-4-space-limit}).
The spectra are modified by the consideration of a sufficiently strong
magnetic field leading also to quantum gravity contributions. The
exact solutions can be derived from string theory. Then adding a magnetic
field to the string theory backgrounds containing a smooth flat $\mathbb{R}^{4}$
results in some curved background by replacing the flat $\mathbb{R}^{4}$
(Subsec. \ref{sub:Marginally-deformed-exact}). In closed string theory,
one has to include the effect of the gravitational field since even
a constant magnetic field carries energy and spacetime has to be curved.
Thus the flat Euclidean 4-space is replaced by the curved $S^{3}\times\mathbb{R}$
geometry. This replacement is rather universal (i.e. independent on
the type of string theory) and results in the appearance of the level
$k$ WZW on $SU(2)$ \cite{KK95}.

From the geometrical point of view the integral class $k\in H^{3}(S^{3},\mathbb{Z})$
can be realized by some exotic smooth $\mathbb{R}_{k}^{4}$ where
the effects of the exotic smoothness can be localized on some $S^{3}\subset\mathbb{R}^{4}$.
This is a kind of a geometric \emph{localization principle} as discussed
in \cite{AsselmeyerKrol2009}. Thus, effects of certain exotic $\mathbb{R}_{k}^{4}$'s,
when localized on $S^{3}$, give rise to the level $k$ WZW on $SU(2)$.
The reason for this coincidence is deeply rooted in the Godbillon-Vey
(GV) class of some codimension one foliations of $S^{3}$. Its evaluation
on the fundamental class of $S^{3}$ is precisely the Wess-Zumino
(WZ) term of the WZW model. Furthermore different GV classes of the
foliation correspond to different exotic $\mathbb{R}_{k}^{4}$ (with
respect to some parameter family of exotic $\mathbb{R}^{4}$'s, the
so-called radial family). This is discussed in the next Section.

The effects of strong magnetic fields and its corresponding gravity
backreactions are obtained as suitable marginal deformations of the
WZW model on $SU(2)$ (Subsec. \ref{sub:Marginally-deformed-exact}
and App. \ref{sec:Appendix-The-deformed-exact}). The string spectra
in the background $SU(2)_{k}\times\mathbb{R}$ depend on these deformations.
In Subsec. \ref{sec:The-energy-spectra} we relate these spectra with
effects caused by exotic $\mathbb{R}_{k}^{4}$'s.Especially we discuss
the large volume limit $k\to\infty$ corresponding to the flat $\mathbb{R}^{4}$
and the rescaling of the $H$ field. Then flat string theory spectra
deformed by $H$ including gravitational effects do not depend on
$k$. However, the exchange of the flat standard $\mathbb{R}^{4}$
by a non-flat exotic $\mathbb{R}_{k}^{4}$ gives a finite $k$ in
the WZW model. Then the dependence of the spectra (see Eqs. (\ref{eq:Mod-4d-spectrum}),
(\ref{eq:Mod-4d-spectrum-2})) on the level $k$ reflects precisely
the effects of the exotic $\mathbb{R}_{k}^{4}$. . 

The whole analysis in this paper used deep mathematical structures
to derive the physical results from some exotic $\mathbb{R}_{k}^{4}$.
As we mentioned above, exotic $\mathbb{R}_{k}^{4}$ are non-flat Riemannian
manifolds, i.e. non-trivial solutions of Einsteins equations \cite{Sladkowski2001}.
As shown here, this effect remains true for the regime where quantum
corrections to gravity become important. Here we found that the structure
of exotic $\mathbb{R}_{k}^{4}$ and its connection to string backgrounds
are the appropriate tools for the calculation of the corrections.

\section{Small exotic $\mathbb{R}_{k}^{4}$, foliations and Wess-Zumino term\label{sec:Small-exotic-,}}

An exotic $\mathbb{R}^{4}$ is a topological space with $\mathbb{R}^{4}-$topology
but with a different (i.e. non-diffeomorphic) smoothness structure
than the standard $\mathbb{R}_{std}^{4}$ getting its differential
structure from the product $\mathbb{R}\times\mathbb{R}\times\mathbb{R}\times\mathbb{R}$.
The exotic $\mathbb{R}^{4}$ is the only Euclidean space $\mathbb{R}^{n}$
with an exotic smoothness structure. The exotic $\mathbb{R}^{4}$
can be constructed in two ways: by the failure to arbitrarily split
a smooth 4-manifold into pieces (large exotic $\mathbb{R}^{4}$) and
by the failure of the so-called smooth h-cobordism theorem (small
exotic $\mathbb{R}^{4}$). Here we deal with the class of small exotic
$\mathbb{R}^{4}$'s which allowing the smooth embedding of a 3-sphere.
In \cite{DeMichFreedman1992}, a family of uncountable many different
small exotic $\mathbb{R}^{4}$ was constructed, the radial family
depending on one parameter, the radius $\rho$. In the following we
will use a countable subfamily $\mathbb{R}_{k}^{4}$ of this family
as explained below.

There are widely discussed difficulties with explicit coordinate descriptions
(see e.g. \cite{Asselmeyer2007}) of different differential structures
on $\mathbb{R}^{4}$ (and on other open 4-manifolds) . In a series
of recent papers, we were however able to relate these 4-exotics with
some other structures on $S^{3}$ (see e.g. \cite{AsselmeyerKrol2009,AsselmeyerKrol2009a,AsselmKrol2011c}).
This 3-sphere $S^{3}$ is supposed to fulfill specific topological
conditions: it has to lie in the ambient $\mathbb{R}^{4}$ as part
of the boundary of some compact 4-submanifold (the so-called Akbulut
cork) with boundary. If all properties are met , one can prove that
exotic smoothness of the $\mathbb{R}^{4}$ is tightly related with
codimension-one foliations of this 3-sphere $S^{3}$, hence with the
3-rd real cohomology classes of $S^{3}$ \cite{AsselmeyerKrol2009}:

\emph{The exotic $\mathbb{R}^{4}$'s from the radial family are determined
by codimension-1 foliations $\mathcal{F}$'s with non-vanishing Godbillon-Vey
(GV) class $GV(\mathcal{F})$ in $H^{3}(S^{3},\mathbb{R})$ of a 3-sphere
lying at the boundary of the Akbulut corks of $\mathbb{R}^{4}$'s.
The radius $\rho$ in the family and the Godbillon-Vey number $GV$as
pairing $\langle GV(\mathcal{F}),[S^{3}]\rangle$ between the GV class
and the fundamental class $[S^{3}]\in H_{3}(S^{3})$ are related by
${\rm GV}=\rho^{2}$. We say: the exoticness is localized at a 3-sphere
inside the small exotic $\mathbb{R}^{4}$ (seen as a submanifold of
$\mathbb{R}^{4}$).}

In the particular case of the integral class as element of \emph{$H^{3}(S^{3},\mathbb{Z})$}
one obtains the relation between the exotic $\mathbb{R}_{k}^{4}$,
$k[\:]\in H^{3}(S^{3},\mathbb{Z})$, $k\in\mathbb{Z}$ and the WZ
term of the $k$ WZW model on $SU(2)$. In \cite{AsselmeyerKrol2009},
we considered particularly this case. Topologically, this case refers
to flat $PSL(2,\mathbb{R})-$bundles over the space $(S^{2}\setminus\left\{ \mbox{\mbox{k} punctures}\right\} )\times S^{1}$
where the gluing of $k$ solid tori produces a 3-sphere (so-called
Heegard decomposition). Then one obtains the relation \cite{AsselmeyerKrol2009}:\begin{equation}
\frac{1}{(4\pi)^{2}}\langle GV(\mathcal{F}),[S^{3}]\rangle=\frac{1}{(4\pi)^{2}}\,\intop_{S^{3}}GV(\mathcal{F})=\pm(2-k)\label{eq:integer-GV}\end{equation}
in dependence on the orientation of the fundamental class $[S^{3}]$.
Now we interpret the Godbillon-Vey invariant as WZ term. For that
purpose we use the group structure $SU(2)=S^{3}$ of the 3-sphere
$S^{3}$ and identify $SU(2)=S^{3}$. Let $g\in SU(2)$ be a unitary
matrix with $\det g=-1$. The left invariant 1-form $g^{-1}dg$ generates
locally the cotangent space connected to the unit. The forms $\omega_{k}=Tr((g^{-1}dg)^{k})$
are complex $k-$forms generating the deRham cohomology of the Lie
group. The cohomology classes of the forms $\omega_{1},\omega_{2}$
vanish and $\omega_{3}\in H^{3}(SU(2),\mathbb{R})$ generates the
cohomology group. Then we obtain for the integral of the generator
\[
\frac{1}{8\pi^{2}}\intop_{S^{3}=SU(2)}\omega_{3}=1\quad.\]
 This integral can be interpreted as winding number of $g$. Now we
consider a smooth map $G:S^{3}\to SU(2)$ with 3-form $\Omega_{3}=Tr((G^{-1}dG)^{3})$
so that the integral\[
\frac{1}{8\pi^{2}}\intop_{S^{3}=SU(2)}\Omega_{3}=\frac{1}{8\pi^{2}}\intop_{S^{3}}Tr((G^{-1}dG)^{3})\in\mathbb{Z}\]
 is the winding number of $G$. Every Godbillon-Vey class with integer
value like (\ref{eq:integer-GV}) is generated by a 3-form $\Omega_{3}$.
Therefore the Godbillon-Vey class is the WZ term of the $SU(2)_{k}$.
Thus we obtain the relation:

\emph{The structure of exotic $\mathbb{R}_{k}^{4}$'s, $k\in\mathbb{Z}$
from the radial family determines the WZ term of the $k$ WZW model
on $SU(2)$.}

Originally this WZ term was introduced to cancel the conformal anomaly
of the classical $\sigma$-model on $SU(2)$. Thus we have a way to
include this cancellation term by using smooth 4-geometry. Then we
can argue that the smoothness of the embedding space $\mathbb{R}^{4}$
of the 3-sphere $S^{3}$ depends on its codimension-1 foliation. In
the exotic case $\mathbb{R}_{k}^{4}$ one obtains the WZ term of the
classical $\sigma$-model with target $S^{3}=SU(2)$. Furthermore
we have the important correlation:

\emph{The change of smoothness from an exotic $\mathbb{R}_{k}^{4}$
to an exotic $\mathbb{R}_{l}^{4}$, $k,\, l\in\mathbb{Z}$ both from
the radial family, corresponds to the change of the level from $k$
to $l$ of the WZW model on $SU(2)$, i.e. $k\,{\rm WZW\to}l\,{\rm WZW}$.}

Let us consider now the end of the exotic $\mathbb{R}_{k}^{4}$ i.e.
$S^{3}\times\mathbb{R}$. This end cannot be standard smooth \cite{GomSti:1999}
and it is in fact an exotic smooth $S^{3}\times_{\Theta_{k}}\mathbb{R}$
\cite{Fre:79}. Using the connection of the exotic $\mathbb{R}_{k}^{4}$
with the WZ term above, we have determined the ,,quantized'' geometry
of $SU(2)_{k}\times\mathbb{R}$ by relating it to the exotic geometry
of the end of $\mathbb{R}_{k}^{4}$. Especially the appearance of
the $SU(2)_{k}\times\mathbb{R}$ is a source for various further constructions.
In particular, the gravitational effects of $\mathbb{R}_{k}^{4}$
on the quantum level are determined via string theory by replacing
consistently the flat $\mathbb{R}^{4}$ part of the background by
the curved 4-space $SU(2)_{k}\times\mathbb{R}$.

\section{Superstring exact solutions and exotic 4-geometries}

We want to use the relation from the previous section and the results
of \cite{AsselmeyerKrol2011,AssKrol2010ICM} to evaluate some physical
effects from the exotic $\mathbb{R}_{k}^{4}$ in the QG regime. In
the following we simply suppose that there is a locally Minkowskian
smooth metric on $\mathbb{R}_{k}^{4}$. 

Exotic $\mathbb{R}_{k}^{4}$ is a smooth Riemannian manifold but its
structure essentially deals with non-commutative geometry and quantization
\cite{AsselmKrol2011c}. The connection with string exact backgrounds
was also recognized \cite{AsselmeyerKrol2011,AssKrol2010ICM}. Especially
with the topological assumptions above (see Sec. \ref{sec:Small-exotic-,})
we obtained the following correlation in our previous work: 
\begin{quotation}
\emph{The change of the smoothness from the standard $\mathbb{R}^{4}$
to exotic $\mathbb{R}_{k}^{4}$, corresponds to the change of exact
string backgrounds from }$\mathbb{R}^{4}\times K^{6}$ to $SU(2)_{k}\times\mathbb{R}_{\phi}\times K^{6}$. 
\end{quotation}
Let us note the importance of the exotic smoothness structure. Otherwise
we are left with separate regimes of smooth 4-geometry (GR) and superstring
theory (QG).

In string theory the above change of backgrounds is one way to include
effects of a strong magnetic field $H$ and its gravitational backreaction
in the 4D part of the background \cite{KK95}. These results show
the existence of a consistent, from the point of view of QG and field
theory, way to change the geometry of flat $\mathbb{R}^{4}$ to the
geometry of a non-flat $SU(2)_{k}\times\mathbb{R}$. The calculations
of correlations functions are performed in heterotic and type II superstring
theories along with superconformal world-sheet symmetry where one
consistently changes between backgrounds (see above) . The coincidence
of the 4D parts of the backgrounds with structures derived from small
exotic $\mathbb{R}_{k}^{4}$ is a tool for the calculation of strong
$H$ QG effects, when $H$ is defined on exotic $\mathbb{R}_{k}^{4}$.
The detailed description of the exact supersymmetric ${\cal N}=4$,
$c=4$ backgrounds (as representations of the superconformal algebra)
can be found in \cite{Antoniadis94,KK95}. In the next section following
\cite{KK95} we discuss the deformations of these backgrounds from
the point of view of $\sigma$-model and CFT constructions (see Appendix
\ref{sec:Appendix-The-deformed-exact}) which represent the effects
of the strong magnetic field $H$ and gravity backreactions on the
backgrounds.

\subsection{Marginally deformed exact string backgrounds\label{sub:Marginally-deformed-exact}}

For a given exotic $\mathbb{R}_{k}^{4}$, $k=1,2,...$, from the radial
family \cite{AsselmeyerKrol2009,AsselmeyerKrol2011} one can consider
the exact background of closed string theory $\mathbb{R}_{\phi}\times SU(2)_{k}\times\mathbb{R}^{5,1}$
\cite{AsselmeyerKrol2011} where the part of the background $\mathbb{R}_{\phi}\times SU(2)_{k}$
appears only when the flat standard $\mathbb{R}^{4}$ is replaced
by the exotic $\mathbb{R}_{k}^{4}$. As shown in \cite{KK95}the replacement
of a flat $\mathbb{R}^{4}$ by $\mathbb{R}_{\phi}\times SU(2)_{k}$
in the string background $\mathbb{R}^{4}\times K^{6}$ is a tool to
include strong (almost) constant magnetic field effects in 4-dimensions
and its corresponding quantum gravitational backreactions. This replacement
was used to calculate the 4-dimensional magneto-gravitational deformations
of the spectra of charged particles\cite{KK95}. We remark that the
change from the flat $\mathbb{R}^{4}$ to the curved 4-dimensional
part: $\mathbb{R}^{4}\times K^{6}\to SU(2)_{k}\times\mathbb{R}_{\phi}\times K^{6}$
is performed certainly under the presence of supersymmetry in 10 dimensions.
Here, we consider the presence of supersymmetry as a technical tool
to effectively perform QG calculations. Firstly, let us recall following
\cite{KK95} how the exact backgrounds and their deformations appear
on the level of $\sigma$-models. We consider the $SO(3)_{k/2}\times\mathbb{R}_{Q}$
CFT case which is the result of a projective map $SU(2)_{k}\times\mathbb{R}_{Q}\to SO(3)_{k/2}\mathbb{R}_{Q}$
for $k$ even. In this case, the action for heterotic $\sigma$-model
is given by:
\begin{widetext}
\begin{equation}
S_{4}=\frac{k}{4}{\rm I}_{SO(3)}(\alpha,\beta,\gamma)+\frac{1}{2\pi}\int d^{2}z\left[\partial x^{0}\overline{\partial}x^{0}+\psi^{0}\partial\psi^{0}+\sum_{a=1}^{3}\psi^{a}\partial\psi^{a}\right]+\frac{Q}{4\pi}\int\sqrt{g}R^{(2)}x^{0}\label{eq:S4action}\end{equation}
\end{widetext}
where ${\rm I}_{SO(3)}(\alpha,\beta,\gamma)=\frac{1}{2\pi}\int d^{2}z\left[\partial\alpha\partial\overline{\alpha}+\partial\beta\partial\overline{\beta}+\partial\gamma\partial\overline{\gamma}+2cos\beta\partial\alpha\partial\gamma\right]$
in the Euler angles of $SU(2)=S^{3}$. Then the bosonic $\sigma$-model
action reads in general:

\begin{equation}
S=\frac{1}{2\pi}\int d^{2}z(G_{\mu\nu}+B_{\mu\nu})\partial x^{\mu}\overline{\partial}x^{\nu}+\frac{1}{4\pi}\int\sqrt{g}R^{(2)}\Phi(x)\end{equation}
 By comparison with (\ref{eq:S4action}) we obtain the non-zero background
fields as:

\begin{equation}
\begin{array}{c}
G_{00}=1,\; G_{\alpha\alpha}=G_{\beta\beta}=G_{\gamma\gamma}=\frac{k}{4},\; G_{\alpha\gamma}=\frac{k}{4}\cos\beta\\[3pt]
B_{\alpha\gamma}=\frac{k}{4}\cos\beta,\;\Phi=Qx^{0}=\frac{x^{0}}{\sqrt{k+2}}\,.\end{array}\end{equation}

Following \cite{PS08}, one can decompose the supersymmetric WZW model
into a bosonic $SU(2)_{k-2}$ part with affine currents $J^{i}$ and
into three free fermions $\psi^{a}$, $a=1,2,3$ in the adjoint representation
of $SU(2)$. Supersymmetry with ${\cal N}=1$ implies for the affine
currents the expression ${\cal J}^{a}=J^{a}-\frac{i}{2}\epsilon^{abc}\psi^{b}\psi^{c}$.
After introducing the complex fermion combinations $\psi^{\pm}=\frac{1}{\sqrt{2}}(\psi^{1}\pm i\psi^{2})$
and the corresponding change of the affine bosonic currents $J^{\pm}=J^{1}\pm iJ^{2}$,
the supersymmetric affine currents read:

\begin{equation}
{\cal J}^{3}=J^{3}+\psi^{+}\psi^{-},\;{\cal J}^{\pm}=J^{\pm}\pm\sqrt{2}\psi^{3}\psi^{\pm}\end{equation}
 Let us redefine the indices in the fermion fields as: $+\to1$, $-\to2$,
then ${\cal J}^{3}=J^{3}+\psi^{1}\psi^{2}$.From the $\sigma$-model
point of view, the vertex for the magnetic field $H$ on the 4-dimensional
curved space $\mathbb{R}_{\phi}\times SU(2)_{k}$ is the exact marginal
operator given by $V_{m}=H(J^{3}+\psi^{1}\psi^{2})\overline{J}^{a}$.
Similarly the vertex for the corresponding gravitational part is $V_{gr}={\cal R}(J^{3}+\psi^{1}\psi^{2})\overline{J}^{3}$,
and represents the really marginal deformations too. The shape of
these operators follow from the fact that the marginal deformations
of the WZW model can be in general constructed as bilinear expressions
in the currents $J$, $\overline{J}$ of the model \cite{Orlando2006}:

\begin{equation}
{\cal O}(z,\overline{z})=\sum_{i,j}c_{ij}J^{i}(z)\overline{J^{j}}(\overline{z})\label{eq:Trully-marginal-deform}\end{equation}
where $J^{i}$, $\overline{J^{j}}$ are left and right-moving affine
currents, respectively \cite{Orlando2006}. Let $\texttt{g}$, $\bar{\texttt{g}}$
be the holomorphic and antiholomorphic Kac-Moody affine algebras of
the $SU(2)_{k}$ WZW model (seen as Lie algebra). The currents $J^{i}$,
$\overline{J^{j}}$ span the abelian Lie sub-algebras $\texttt{h}$,
$\bar{\texttt{h}}$ of the Lie algebras $\texttt{g}$, $\bar{\texttt{g}}$.
Then the Lie groups $U(1)^{d}$, $U(1)^{\bar{d}}$ with $d={\rm dim}{\rm \texttt{h}}$
and $\bar{d}={\rm dim}{\rm \bar{\texttt{h}}}$ are associated to the
Lie algebras $\texttt{h}$, $\bar{\texttt{h}}$. The complete class
of the marginal deformations of the WZW model (\ref{eq:Trully-marginal-deform})
are determined by $\frac{O(d,d)}{O(d)\times O(d)}$ interpreted as
transformations of the lattice of the charges (\cite{Orlando2006},
p. 20). In case of the single $U(1)$ subalgebra (and a single deforming
field) we have $O(1,1)$ deformations (see Appendix \ref{sec:Appendix-The-deformed-exact}).

Here, following \cite{KK95}, we consider a covariantly constant magnetic
field $H_{i}^{a}=\epsilon^{ijk}F_{jk}^{a}$ and a constant curvature
${\cal R}^{il}=\epsilon^{ijk}\epsilon^{lmn}{\cal R}_{jmkn}$ in the
4-dimensional background in the closed superstring theory as discussed
above. When this field is in the $\mu=3$ direction the deformation
is proportional to $(J^{3}+\psi^{1}\psi^{2})\overline{J}$ and the
right moving current $\overline{J}$ is normalized as $<\overline{J}(1)\overline{J}(0)>=k_{g}/2$.
Rewriting the currents in Euler angles, i.e. $J^{3}=k(\partial\gamma+\cos\beta\partial\alpha)$,
$\overline{J}^{3}=k(\overline{\partial}\alpha+\cos\beta\overline{\partial}\gamma)$,
we get for the perturbation of the (heterotic) action in (\ref{eq:S4action}),
the following expression:

\begin{equation}
\delta S_{4}=\frac{\sqrt{kk_{g}}H}{2\pi}\int d^{2}z(\partial\gamma+\cos\beta\partial\alpha)\overline{J}\:.\label{eq:dS4}\end{equation}
 The new $\sigma$-model with action $S_{4}+\delta S_{4}$ is again
conformally invariant for all orders in $\alpha'$ since:
\begin{widetext}
\[
S_{4}+\delta S_{4}=\frac{k}{4}{\rm I}_{SO(3)}(\alpha,\beta,\gamma)+\delta S_{4}+\frac{k_{g}}{4\pi}\int d^{2}z\partial\phi\overline{\partial}\phi=\frac{k}{4}{\rm I}_{SO(3)}(\alpha,\beta,\gamma+2\sqrt{\frac{k_{g}}{k}}H\phi)+\frac{k_{g}(1-2H^{2})}{4\pi}\int d^{2}z\partial\phi\overline{\partial}\phi\:.\]
\end{widetext}
This shows the marginal character of the magnetic deformation where
we have chosen for the currents $J$ and $\overline{J}$ their bosonizations
$\partial\phi$ and $\overline{\partial}\phi$ correspondingly.

To the perturbation (\ref{eq:dS4}) there is a corresponding background
determined by background fields like a graviton $G_{\mu\nu}$, gauge
fields $F_{\mu\nu}^{a}$, an antisymmetric field (three form) $H_{\mu\nu\rho}$
and a dilaton $\Phi$, which, in turn, are solutions to the following
equations of motion:

\begin{equation}
\begin{array}{c}
\frac{3}{2}\left[4(\nabla\Phi)^{2}-\frac{10}{3}\square\Phi-\frac{2}{3}R+\frac{1}{12g^{2}}F_{\mu\nu}^{a}F^{a,\mu\nu}\right]=C\\[4pt]
R_{\mu\nu}-\frac{1}{4}H_{\mu\nu}^{2}-\frac{1}{2g^{2}}F_{\mu\rho}^{a}F_{\nu}^{a\rho}+2\nabla_{\mu}\nabla_{\nu}\Phi=0\\[4pt]
\nabla^{\mu}\left[e^{-2\Phi}H_{\mu\nu\rho}\right]=0\\[4pt]
\nabla^{\nu}\left[e^{-2\Phi}F_{\mu\nu}^{a}\right]-\frac{1}{2}F^{a,\nu\rho}H_{\mu\nu\rho}e^{-2\Phi}=0\;\end{array}\label{EOM}
\end{equation}
These equations can be derived by the variations of the following
effective 4-dimensional gauge theory action:
\begin{equation}
\begin{array}{c}

S=\int d^{4}x\sqrt{G}e^{-2\Phi}[R+4(\nabla\Phi)^{2}-\frac{1}{12}H^{2}- \\ [4pt]
-\frac{1}{4g^{2}}F_{\mu\nu}^{a}F^{a,\mu\nu}+\frac{C}{3}]
\end{array}
\end{equation}
where $C$ is the l.h.s. of the first equation in (\ref{EOM}). Here
we set $g_{str}=1$ and for the gauge coupling $g^{2}=2/k_{g}$ The
fields $F_{\mu\nu}^{a},H_{\mu\nu\rho}$ are usually defined by $F_{\mu\nu}^{a}=\partial_{\mu}A_{\nu}-\partial_{\nu}A_{\mu}+f^{abc}A_{\mu}^{b}A_{\nu}^{c}$,
$H_{\mu\nu\rho}=\partial_{\mu}B_{\nu\rho}-\frac{1}{2g^{2}}\left[A_{\mu}^{a}F_{\nu\rho}^{a}-\frac{1}{3}f^{abc}A_{\mu}^{a}A_{\nu}^{b}A_{\rho}^{c}\right]+{\rm permutations}$.
The constants $f^{abc}$ are structure constants of the gauge group
and $A_{\mu}^{a}$ is the effective gauge field. One can notice that
the term in $H_{\mu\nu\rho}$ enclosed by square brackets is the Chern-Simons
term for the gauge potential $A_{\mu}^{a}$.Now a solution of these
equations for the background agreeing with the deformation (\ref{eq:dS4}),
reads:

\begin{equation}
\begin{array}{c}
G_{00}=1,\: G_{\beta\beta}=\frac{k}{4}\,,\; G_{\alpha\gamma}=\frac{k}{4}(1-2H^{2})\cos\beta\\[4pt]
G_{\alpha\alpha}=\frac{k}{4}(1-2H^{2}\cos^{2}\beta)\,,\; G_{\gamma\gamma}=\frac{k}{4}(1-2H^{2})\,,\\[4pt]
B_{\alpha\gamma}=\frac{k}{4}\cos\beta\\[4pt]
A_{\alpha}=g\sqrt{k}H\cos\beta\,,\; A_{\gamma}=g\sqrt{k}H\,,\;\Phi=\frac{x^{0}}{\sqrt{k+2}}\,.\end{array}\label{BCKG-H}\end{equation}
where $H$ is the magnetic field as in (\ref{eq:dS4}).

Similarly, when gravitational marginal deformations like in the vertex
$V_{gr}={\cal R}(J^{3}+\psi^{1}\psi^{2})\overline{J}^{3}$ are included
one can derive a corresponding exact background of string theory by
$\sigma$-model calculations \cite{SenHass93,KK95}. Again, one obtains
for the fields in this background by solving the effective field theory
equations (\ref{EOM})\cite{KK95}:

\begin{equation}
\begin{array}{c}
G_{00}=1,\: G_{\beta\beta}=\frac{k}{4}\\[4pt]
G_{\alpha\alpha}=\frac{k}{4}\frac{(\lambda^{2}+1)^{2}-(8H^{2}\lambda^{2}+(\lambda^{2}-1)^{2})\cos^{2}\beta)}{(\lambda^{2}+1+(\lambda^{2}-1)\cos\beta)^{2}}\\[4pt]
G_{\gamma\gamma}=\frac{k}{4}\frac{(\lambda^{2}+1)^{2}-(8H^{2}\lambda^{2}-(\lambda^{2}-1)^{2})\cos^{2}\beta)}{(\lambda^{2}+1+(\lambda^{2}-1)\cos\beta)^{2}}\\[4pt]
G_{\alpha\gamma}=\frac{k}{4}\frac{4\lambda^{2}(1-2H^{2})\cos\beta+(\lambda^{4}-1)\sin^{2}\beta}{(\lambda^{2}+1+(\lambda^{2}-1)\cos\beta)^{2}}\\[4pt]
B_{\alpha\gamma}=\frac{k}{4}\frac{\lambda^{2}-1+(\lambda^{2}+1)\cos\beta}{(\lambda^{2}+1+(\lambda^{2}-1)\cos\beta)^{2}}\\[4pt]
A_{\alpha}=2g\sqrt{k}\frac{H\lambda\cos\beta}{(\lambda^{2}+1+(\lambda^{2}-1)\cos\beta)^{2}}\\[4pt]
A_{\gamma}=2g\sqrt{k}\frac{H\lambda}{(\lambda^{2}+1+(\lambda^{2}-1)\cos\beta)^{2}}\\[4pt]
\Phi=\frac{t}{\sqrt{k+2}}-\frac{1}{2}\log\left[\lambda+\frac{1}{\lambda}+(\lambda-\frac{1}{\lambda})\cos\beta\right]\end{array}\label{BCKG-G-H}\end{equation}
We remark that the dependence on $\lambda$ (see App. \ref{sec:Appendix-The-deformed-exact})
shows the existence of gravitational backreaction which was absent
in the purely magnetic deformed background (\ref{BCKG-H}).

\subsection{The mass gap in the background with linear dilaton\label{sub:The-mass-gap}}

In field theory a dilaton $\Phi$ couples to a massless bosonic field
$T$ in a universal fashion: \begin{equation}
S[\Phi,T]=\int e^{-2\Phi}\partial_{M}T\partial^{M}T.\label{eq:Dilaton-Mass-Gap-1}\end{equation}
 which can be rewritten by introducing a new field $U=e^{-\Phi}T$
to:

\begin{equation}
S[\Phi,U]=\int\partial_{M}U\partial^{M}U+[\partial^{2}\Phi-\partial_{M}\Phi\partial^{M}\Phi]U\label{eq:Dilaton-Mass-Gap-2}\end{equation}
 Thus for a linear dilaton $\Phi=q_{M}X^{M}$ the field $U$ becomes
massive with mass square $M^{2}=q_{M}q^{M}$ (for $X^{M}$ spacelike).
In this way the massless boson $T$ is mapped to the boson $U$ with
the mass $M$. However this mechanism does not work in case of massless
free fermions. But in a four dimensional spacetime, the chiral fermion
$\psi$ can be coupled to an antisymmetric tensor $H_{\mu\nu\rho}$
like: \[
S[\psi,H]=\int\overline{\psi}\gamma^{\mu}\left[\overleftrightarrow{\partial_{\mu}}+H_{\mu}\right]\psi\]
 where $H_{\mu}=\varepsilon^{\mu\nu\rho\sigma}H_{\nu\rho\sigma}$
is the dual of the antisymmetric tensor $H_{\nu\rho\sigma}$. This
system can be embedded into a string background with the fields: $\Phi$
and $H_{MNP}$. Then by using one-loop string equations: \begin{equation}
\begin{array}{c}
R_{MN}=-2\nabla_{M}\nabla_{N}\Phi+\frac{1}{4}H_{MPR}H_{N}^{\text{ \ \ }PR},\\[4pt]
\nabla_{L}\left(e^{-2\Phi}H_{MN}^{L}\right)=0,\\[4pt]
\nabla^{2}\Phi-2\left(\nabla\Phi\right)^{2}=-\frac{1}{12}H^{2},\end{array}\label{String-one-loop-1}\end{equation}
one gets for the linear dilaton $\Phi=q_{M}X^{M}$ the following relation:
\[
q_{M}q^{M}=\frac{1}{6}H^{2}\]
 and the scalar curvature $R$ is: \[
R=\frac{3}{2}q_{M}q^{M}\,.\]
 If non-vanishing components of $q_{M},X^{M}$ and $H_{MNP}$ are
in the direction of the four dimensional space one obtains: \[
q^{\mu}\sim\varepsilon^{\mu\nu\rho\sigma}H_{\nu\rho\sigma}.\]
Thus the Dirac operator acquires a mass gap proportional to $q_{\mu}q^{\mu}$.

The problem to embedd a four dimensional fermion system in the exact
string background was considered in \cite{KK95b}. 

Now we consider the case when the four dimensional space is represented
by the $\mathbb{R}_{\phi}\times SU(2)_{k}$ part of the string background.
Then the linear dilaton is given by $\Phi=QX^{0}$ where $Q$ is related
to the level $k$ of the WZW model on $SU\left(2\right)$ by $Q=\left(k+2\right)^{-1/2}$.
Therefore the corresponding CFT has the same central charge like in
the flat case. Hence the direct consequence of (\ref{eq:Dilaton-Mass-Gap-2})
is that the massless bosons acquire the mass gap $\Delta M^{2}=\mu^{2}=\left(k+2\right)^{-1}$.

Given the exact string background with four dimensional part $\mathbb{R}_{\phi}\times SU(2)_{k}$,
we consider the perturbation of this theory by gauge and gravitational
fields. Now we will analyze the existence of the mass gap by some
explicit calculations in the perturbed theory. . 

The minimal value of the mass of bosonic states in the deformed theory
reads:

\begin{equation}
\begin{array}{c}
M_{\min}^{2}\left(\mathcal{Q},I\right)=\frac{1}{2}\left(\mathcal{Q}^{2}-1\right)+\frac{1}{k+2}\left[\left(|I|+1/2\right)^{2}-\left(\mathcal{Q}+I\right)^{2}\right]+\\[4pt]
\frac{1}{2}\left(1+\sqrt{1+F^{2}}\right)\left(\frac{\mathcal{Q}+I}{\sqrt{k+2}}+eH\right)^{2},\end{array}\label{String-2}\end{equation}
where $|\mathcal{Q}|=1$ and $|I|=0,1,...,k/2$ and $\mathcal{Q},I,{\cal P}$
are the zero-modes of the corresponding currents in this theory (see
\cite{KK95} and Appendix \ref{sec:Appendix-The-deformed-exact}).
We obtain for the magnetic field $H=F/\left(\sqrt{2}\left(1+\sqrt{1+F^{2}}\right)\right)$
and the electric charge $e=\sqrt{2}\overline{\mathcal{P}}/\sqrt{k_{g}}$.
In the discussed case the function $F$ is given by:\[
F=\sinh\left(2x\right),\]
 where $x$ is a parameter of the group $O\left(1,1\right)$. Thus
the magnetic field $H$ takes the form: \[
H=\frac{1}{\sqrt{2}}\tanh x.\]
 For $x=0$ the minimal value of the mass is:\[
M_{\min}^{2}\left(\mathcal{Q},I\right)=\frac{1}{2}\left(\mathcal{Q}^{2}-1\right)+\frac{\left(|I|+1/2\right)^{2}}{k+2}.\]
 Since $|\mathcal{Q}|=1$ the first term vanishes and one gets: \[
M_{\min}^{2}\left(\mathcal{Q},I\right)=\frac{\left(|I|+1/2\right)^{2}}{k+2}.\]
 This value is proportional to the mass gap obtained from the linear
dilaton.

Now let us compute the following difference of mass squares: $\Delta M_{\left(1\right)}^{2}=M_{\min}^{2}\left(+1,I\right)-M_{\min}^{2}\left(-1,I\right)$:
\[
\Delta M_{\left(1\right)}^{2}=\frac{4}{k+2}S\left(I\right),\]
 where the function $S\left(I\right)$ is equal to: \[
S\left(I\right)=2I\left(-1+\sqrt{1+F^{2}}\right)+\frac{\sqrt{2}}{4}eF\sqrt{k+2}.\]

\[
\Delta M_{\left(1\right)}^{2}=\frac{16I\sinh^{2}x}{k+2}+\frac{\sqrt{2}}{4}\frac{e\sinh\left(2x\right)}{\sqrt{k+2}}.\]
 For $x=0$ (the magnetic perturbation is switched off at this stage
of the theory) one gets: $\Delta M_{\left(1\right)}^{2}=0$, hence
the mass is minimal for any $\mathcal{Q}=\pm1$. Next we consider
another difference of square masses of the string states: $\Delta M_{\left(2\right)}^{2}=M_{\min}^{2}\left(\mathcal{Q},I+1\right)-M_{\min}^{2}\left(\mathcal{Q},I\right)$.
In this case we get:

\begin{equation}
\begin{array}{c}
\Delta M_{\left(2\right)}^{2}\left(I;x\right)=\\[4pt]
\frac{1}{k+2}\left[2I+|I+1|-|I|+1+\left(2\mathcal{Q}+2I+1\right)\sinh^{2}x\right]+
\frac{e\sinh\left(2x\right)}{\sqrt{2}\sqrt{k+2}}.\end{array}\label{eq: String-3}\end{equation}
For $x=0$ we get:\[
\Delta M_{\left(2\right)}^{2}\left(I;0\right)=\frac{1}{k+2}\left[2I+|I+1|-|I|+1\right].\]
 Because: \[
2I+|I+1|-|I|+1=\left\{ \begin{array}{ccc}
2I+2 & for & I\geq0\\
4I+2 & for & I\in[-1,0]\\
2I & for & I\leq-1\end{array}\right.\]
we obtain:

\[
\Delta M_{\left(2\right)}^{2}\left(I;0\right)=\left\{ \begin{array}{ccc}
\frac{2I+2}{k+2} & for & I\geq0\\
\frac{2I}{k+2} & for & I\leq-1\end{array}\right.\]
So again linear dilaton mass gap is smaller than this difference.

\section{The energy spectra in dimension 4\label{sec:The-energy-spectra}}

\subsection{Field theory vs. string theory spectra of charged particles in standard
4-space\label{sub:Flat-4-space-limit}}

In case of the magnetic field on $S^{3}$ we choose for the vector
potential $A_{\mu}$ in the model (\ref{BCKG-H}):

\begin{equation}
A_{\alpha}=H\cos\beta\,,\; A_{\beta}=0\,,\; A_{\gamma}=H\,.\end{equation}
 The Hamiltonian for a particle of electric charge $e$ moving on
$S^{3}$ is given by

\begin{equation}
\overline{\textbf{H}}=\frac{1}{\sqrt{{\rm det}G}}(\partial_{\mu}-ieA_{\mu})\sqrt{{\rm det}G}G^{\mu\nu}(\partial_{\nu}-ieA_{\nu})\:.\label{eq:H-on-S3}\end{equation}
 where we assume the standard metric $G_{\mu\nu}$ on $S^{3}$. The
energy spectrum for $\overline{\textbf{H}}$ is then given by: \begin{equation}
\Delta E_{j,m}=\frac{1}{R^{2}}\left[j(j+1)-m^{2}+(eH-m)^{2}\right]\label{eq:E-spectrum-on-S3}\end{equation}
 where $j\in\mathbb{Z}$ and $-j\leq m\leq j$ (like in the case of
the group $SO(3)$). In the flat limit we retrieve the Landau spectrum
of spinless particles in the 3-dimensional space :

\begin{equation}
\Delta E_{n,p_{3}}=e\tilde{H}(2n+1)+p_{3}^{2}+{\cal O}(R^{-1})\end{equation}
 where we rescale $eH$ to $eH=e\tilde{H}+\kappa R+{\cal O}(1)$ and
$m=e\tilde{H}R^{2}+(p_{3}+\kappa)+{\cal O}(1)$. This behavior can
be derived by rewriting the spectrum (\ref{eq:E-spectrum-on-S3})
by using the new parameter $n$: $j=|m|+n$ for $|m|,$$n\in\mathbb{N}$
into $\Delta E_{n,m}=\frac{1}{R^{2}}\left[n(n+1)+|m|(2n+1)\right]+\left(\frac{eH-m^{2}}{R}\right)^{2}$
.

Let us, again following \cite{KK95}, calculate the spectrum in case
of the exact string background (\ref{BCKG-H}). If one choose the
metric (\ref{BCKG-H}) of the background then one is able to derive
the eigenvalues of the Hamiltonian (\ref{eq:E-spectrum-on-S3}) with
the result \cite{KK95}:

\begin{equation}
\Delta E_{j,m}=\frac{1}{R^{2}}\left[j(j+1)-m^{2}+\frac{(eHR-m)^{2}}{(1-2H^{2})}\right]\:.\label{eq:E-spec-in-strings-H}\end{equation}
 With the abbreviations $n\in\mathbb{N}$ by $j=|m|+n$, $|m|=0,\,1/2,\,1,\,...$
we can rewrite the spectrum (\ref{eq:E-spec-in-strings-H}) as:

\begin{equation}
\Delta E_{n,m}=\frac{1}{R^{2}}[n(n+1)+|m|(2n+1)]+\left(\frac{eHR-m}{R\sqrt{1-2H^{2}}}\right)^{2}\label{eq:En,m}\end{equation}
 where the energy spectrum contains the corrections due to the $H$
field appearing in the exact string background (\ref{BCKG-H}). But
we remark that the Hamiltonian (\ref{eq:H-on-S3}) is the Hamiltonian
of a 4-dimensional theory.

Using \cite{KK95,KK95c} and the appendix \ref{sec:Appendix-The-deformed-exact},
one can also calculate the energy spectrum in this background in comparison
to (\ref{eq:En,m}). Importantly, we are able to get from one spectrum
to the other by the following rules:

\begin{equation}
\begin{array}{c}
R^{2}\to k+2\,,\; m\to{\cal Q}+J^{3}\,,\; e\to\sqrt{\frac{2}{k_{g}}}\overline{{\cal Q}}\\[3pt]
H\to\frac{F}{\sqrt{2}(1+\sqrt{1+F^{2}})}=\frac{1}{2\sqrt{2}}\left[F-\frac{F^{3}}{4}+{\cal O}(F^{5})\right]\end{array}\label{Dic1}\end{equation}
where \[
F^{2}=\left\langle F_{\mu\nu}^{a}F_{a}^{\mu\nu}\right\rangle =\intop_{SU(2)}F_{\mu\nu}^{a}F_{a}^{\mu\nu}dvol(SU(2))\]
 is the integrated (squared) field strength with $H_{i}^{a}=\epsilon^{ijk}F_{jk}^{a}$
.

For a particle with spin $S$ (setting $S=Q$) we obtain the following
modifications of the spectrum due to the above rules\cite{KK95}:

\begin{equation}
\Delta E_{j,m,S}=\frac{1}{k+2}\left[j(j+1)-(m+S)^{2}+\frac{(eHR-m-S)^{2}}{(1-2H^{2})}\right]\:.\label{eq:Mod-4d-spectrum-2}\end{equation}

The next step is the inclusion of the gravitational backreactions.
One starts with the string background (\ref{BCKG-G-H}) and computes
again the eigenvalues of (\ref{eq:H-on-S3}). The result for scalar
particles is \cite{KK95}:
\begin{widetext}
\begin{equation}
\Delta E_{j,m,\overline{m}}=\frac{1}{R^{2}}\left[j(j+1)-m^{2}+\frac{(2ReH-(\lambda+\frac{1}{\lambda})m-(\lambda-\frac{1}{\lambda})\overline{m})^{2}}{4(1-2H^{2})}\right]\label{eq:Spectrum G+H}\end{equation}
\end{widetext}
 where $-j\leq m,\overline{m}\leq j$. Again, we obtain for even $k$
(see (\ref{Ap5}) in the Appendix \ref{sec:Appendix-The-deformed-exact})
the corresponding rules in the case where gravity backreactions are
included:

\begin{equation}
\begin{array}{c}
R^{2}\to k+2\,,\; m\to{\cal Q}+J^{3}\,,\; e\to\sqrt{\frac{2}{k_{g}}}\overline{{\cal P}}\,,\;\overline{m}\to\overline{J}^{3}\\[5pt]
H^{2}\to\frac{1}{2}\frac{F^{2}}{F^{2}+2(1+\sqrt{1+F^{2}+{\cal R}^{2}})}\,,\;\lambda^{2}=\frac{1+\sqrt{1+F^{2}+{\cal R}^{2}}+{\cal R}}{1+\sqrt{1+F^{2}+{\cal R}^{2}}-{\cal R}}\,\end{array}\label{Dic2}\end{equation}
 where \[
{\cal R}^{2}=\left\langle R_{\mu\nu\rho\sigma}R^{\mu\nu\rho\sigma}\right\rangle =\intop_{SU(2)}R_{\mu\nu\rho\sigma}R^{\mu\nu\rho\sigma}dvol(SU(2))\]
is the integrated squared scalar curvature and $R_{\mu\nu\rho\sigma}$
is the Riemann tensor of the ,,squashed'' $SU(2)=S^{3}$ in the deformed
background.

\subsection{Exotic 4-geometry limit}

In the previous sections we obtained a dictionary (\ref{Dic2}) to
modify the spectrum (\ref{eq:Spectrum G+H}) of a scalar particle
with charge $e$ by the influence of a magnetic field $H$ and the
inclusion of the gravitational backreaction. Then the dependence on
the even level $k$ emerges:
\begin{widetext}
\begin{equation}
\Delta E_{j,m,\overline{m}}^{k}=\frac{1}{k+2}[j(j+1)-m^{2}]+\frac{(2\sqrt{k+2}eH-(\lambda+\frac{1}{\lambda})m-(\lambda-\frac{1}{\lambda})\sqrt{(1+2/k)}\overline{m})^{2}}{4(k+2)(1-2H^{2})}\:.\label{eq:Mod-4d-spectrum}\end{equation}
\end{widetext}
 Based on our discussion above and in Sec. \ref{sec:Small-exotic-,}
we consider (\ref{eq:Mod-4d-spectrum}) as the modification of the
spectrum of a scalar quantum particle with charge when moving through
a 4-region with exotic geometry $\mathbb{R}_{k}^{4}$ where the scaling
of $H$ is required on the exotic background. Even though we do not
determine explicitly this scaling (this would require the unknown
exotic functions on $\mathbb{R}_{k}^{4}$) the spectrum is exactly
derived. Moreover, the local description of an exotic 4-space remains
similar to the standard case.

Now we want to comment on the possible physics behind the emergence
of these quantum gravitational effects induced by an exotic $\mathbb{R}_{k}^{4}$
in spacetime. Each exotic $\mathbb{R}_{k}^{4}$, $k=1,2,...$ is a
member of the radial family of exotic $\mathbb{R}^{4}$ in the standard
$\mathbb{R}^{4}$. This means that exotic smoothness of $\mathbb{R}_{k}^{4}$
is confined to the open subset of the radii $\sim\sqrt{k}$ \cite{AsselmeyerKrol2009}.
However this exotic smooth 4-region cannot be extended smoothly (with
respect to the standard coordinates) over bigger area and larger radii
or cannot be smoothly glued into the bigger standard $\mathbb{R}^{4}$.
Let us consider the exotic structure of $\mathbb{R}_{k}^{4}$ in the
,,infinite'' extendible $\mathbb{R}^{4}$ denoted by $\check{\mathbb{R}}_{k}^{4}\subset\mathbb{R}^{4}$.
From the point of view of standard smoothness on $\mathbb{R}^{4}$,
the exotic smoothness structure of $\check{\mathbb{R}}_{k}^{4}$ can
be interpreted as a matter sources for gravity. Let us suppose that
there is a definite density of gravitational energy in this $\check{\mathbb{R}}_{k}^{4}$.
Next let us shrink this $\check{\mathbb{R}}_{k}^{4}$ to the region
of a finite diameter $\sim h_{d}(\sqrt{k})$ in the standard 4-space.
\footnote{This standard smooth contraction is however not smooth from the point
of view of $\check{\mathbb{R}}_{k}^{4}$.%
} The amount of energy is contracted and confined in a smaller bounded
4-region and for some, appropriately small diameter, quantum gravity
effects become important and dominate. In the same time every exotic
$\mathbb{R}_{k}^{4}$ determines the WZ term of the $SU(2)_{k}$ WZW
model and the geometry of the end determines the $SU(2)_{k}\times\mathbb{R}$
via its smoothness structure (see Sec. \ref{sec:Small-exotic-,} and
\cite{AsselmeyerKrol2009}). When a particle moves through such exotic,
contracted, 4-region the QG effects dominates and are calculated via
$SU(2)_{k}$ WZW model as in string theory on these backgrounds.

When the radii $\sqrt{k}$ of the $SU(2)$ (seen as 3-sphere) is assigned
to the size of the 4-region in spacetime , one has the dominance of
the classical 4-geometry for large radii. A particle moves via exotic
smooth trajectories determined with respect to the Riemannian geometry
of the non-contracted $\mathbb{R}_{k}^{4}$. The QG effects are calculated
via models of 2-d CFT hence they are preserved under the 2-d scaling
of coordinates (conformal transformations). This means that the non-contracted
$\check{\mathbb{R}}_{k}^{4}$ contains also some QG ingredients but
these are dominated by classical Riemannian 4-geometry and are negligible.

Let us consider the change of the smooth structures $\mathbb{R}_{k}^{4}$
with the size of the region $\sqrt{k}$ in $\mathbb{R}^{4}$, then
the non-contracted limit corresponds to the flat $\mathbb{R}^{4}$
and this is precisely the limit of large $k$ of $SU(2)_{k}$ WZW.
The classical limit of 4-d QG is thus achieved when $k\to\infty$
and in this limit, in the case of Eq. (\ref{eq:Mod-4d-spectrum-2})
(without gravitational backreactions from $H$), there appears the
continuum spectrum. Thus when we shift the standard 4-smoothness of
the $\mathbb{R}^{4}$ into an exotic $\mathbb{R}_{k}^{4}$ and confine
it into the 4-region of diameter $\sim\sqrt{k}$,\emph{ the energy
spectrum of a charged particle moving through the exotic region, without
inclusion of gravitational backreactions from the magnetic field,
becomes quantized}.\emph{ Moreover, the mass gap in the matter spectrum
appears,} $\mu\sim\frac{1}{\sqrt{k+2}}$, which disappears in the
flat $\mathbb{R}^{4}$ limit (Sec. \ref{sub:The-mass-gap}). In the
,,classical'' gravity (flat) limit $k\to\infty$, the spectrum (\ref{eq:Mod-4d-spectrum})
reflects some modified Landau levels including gravitational backreactions.
This spectrum is further modified as in (\ref{eq:Mod-4d-spectrum})
by the presence of an exotic $\mathbb{R}_{k}^{4}$. Again, the mass
gap appears due to the exotic 4-smoothness underlying the 4D spacetime.

From the interpretation above we got a direct indication of quantum
gravitational effects derived from the non-standard smooth metrics
on $\mathbb{R}_{k}^{4}$. More sophisticated connections with QFT
and quantization were worked out in \cite{AsselmKrol2011c} containing
deep topological and differential-geometric results.

The scaled field $H$ is only \emph{locally} constant on the exotic
$\mathbb{R}_{k}^{4}$. Then gravitational backreactions from the field
$H$ are derived as the quantum response of the exotic background
$\mathbb{R}_{k}^{4}$ due to its Riemannian curvature, i.e. the density
of gravitational energy. On the contracted exotic $\mathbb{R}_{k}^{4}$,
however, there dominates the quantum gravity component which deform
consistently the spectra of quantum particles. Taking the exotic $\mathbb{R}_{k}^{4}$
limit (finite $k$) instead of flat $\mathbb{R}^{4}$ ($k\to\infty$)
explains the dependence on $k$. Fixing $k$ in (\ref{eq:Mod-4d-spectrum})
refers to the amount of gravity which is contained in the geometry
of the $\mathbb{R}_{k}^{4}$, the quantum numbers $j$, $m$, $\overline{m}$
are related with symmetries of the (contracted) exotic $\mathbb{R}_{k}^{4}$,
the gravitational backreactions of $H$ are encoded in the dependence
on the modulus $\lambda$.

However, this kind of overlapping between \emph{geometry, quantum
matter }and\emph{ gravity} predicts in fact a new way in which QG
is rooted in 4-dimensions and how it interacts with matter and geometry.
Namely, let us recall that GR predicts the link between the geometry
of spacetime and matter-energy fields by the Einstein field equations
$G_{\mu\nu}=8\pi T_{\mu\nu}$ where $G_{\mu\nu}$ is now the Einstein
tensor and $T_{\mu\nu}$ the energy-momentum tensor. Quantum fields
carry energy and mass i.e. according to GR they disturb the curvature
of spacetime but the geometry affects the propagation of quantum matter
fields. In the presented paper, we have calculated the influence of
the smooth (exotic) 4-geometry on the quantum particles and field
$H$ in the 4-dimensional spacetime. The calculations went through
the quantum level of gravity as in 10d superstring theory and dominated
the classical smooth Riemannian 4-geometry. This scenario shows that
quantizing the gravity field and quantum matter in dimension 4 can
be exactly solved via non-standard smoothness on $\mathbb{R}^{4}$,
where the necessary tools are 10 dimensional. The particular case
presented in the paper has a more universal meaning: it uncovers a
new way of interaction of gravity and quantum matter (see also \cite{AsselmKrol11}).
We will further work on this important issue (cf. \cite{Asselmeyer-Maluga2010,Krol2010,Krol2010b,AsselmKrol11}).

\section{Summary}

In this paper we presented the influence of the exotic $\mathbb{R}_{k}^{4}$
on some energy spectra of charged particles in spacetime whose smoothness
is defined with respect to this exotic $\mathbb{R}_{k}^{4}$. In our
model, the effects are exactly calculable via techniques of superstring
theory in curved backgrounds and 2d CFT. The whole procedure shows
that the Riemannian geometry of exotic $\mathbb{R}_{k}^{4}$ emerges
in the correct classical limit for the quantum gravity represented
by superstring theory at least in the case of the $SU(2)_{k}\times\mathbb{R}_{\phi}\times K^{6}$
backgrounds. Why is superstring theory well suited for exotic $\mathbb{R}_{k}^{4}$
effects in 4-dimensions? As the proposed model shows superstring theory
is a universal and consistent way to describe the change of the 4-dimensional
part of the background from the flat $\mathbb{R}^{4}$ to the curved
$SU(2)_{k}\times\mathbb{R}_{\phi}$. The supersymmetric theory on
these backgrounds is minimally perturbed compared to the flat one
\cite{KK95}, and the exact results, such as QG corrections, are derivable.
Exactly this change of 4-geometry, the shift from the flat 4-space
to the exotic $\mathbb{R}_{k}^{4}$ was derived from topological considerations
. Thus we have a way to deal with the quantum effects of gravity which
is confined to exotic $\mathbb{R}_{k}^{4}$. Superstring theory is
UV finite and unifies gravity with other interactions which takes
place in large energies ${\cal O}(M_{str})\simeq10^{17}$ GeV. However,
the excited string states become important for these energies and
effective string field theory has to be modified. The results in the
paper show that a modified 4-dimensional field theory on some exotic
$\mathbb{R}_{k}^{4}$ should be considered in this regime.

Hence, this $\mathbb{R}_{k}^{4}$ Riemannian geometry becomes an important
ingredient of the classical limit for superstring theory so that 4-dimensional
field theory emerging in the classical limit with well defined QG
corrections should be formulated on spacetimes with exotic 4-geometry
(cf. \cite{AsselmKrol2011c}). The appearance of the mass gap in the
matter spectrum due to nonstandard smoothness is an indication in
favor of this point of view. Moreover the curvature of the string
background considered here (hence the mass gap) can act as the infrared
regulator for string and field theory (see e.g. \cite{KK95a}), so
exotic $\mathbb{R}_{k}^{4}$ can play a role as a natural 4-geometric
regulator. These interesting points we will address in our forthcoming
work.

The results in the paper show that gravity (classical and quantum),
geometry and quantum fields can be related with each other in a different
way than usually prescribed by QFT on semi-classical backgrounds or
by various techniques from superstring theory. The fact that superstring
theory is UV finite and is the theory of quantum gravity containing
other interactions is crucial for the results. The entire approach
works due to the existence of non-standard geometries exclusively
in dimension 4 which appear as new and important ingredients of a
(final) theory of 4-dimensional QG.

\appendix
\section{The deformed exact string spectra\label{sec:Appendix-The-deformed-exact}}

Following \cite{KK95,KK95c} we present now the CFT calculations leading
to the description of the string backgrounds deformed by magnetic
and corresponding gravitational marginal deformations. In the case
of a single magnetic field $F$ the operators corresponding to truly
marginal deformations and in the case of the current-current interactions,
are given by the bilinear product of currents (\ref{eq:Trully-marginal-deform}):
$V_{F}=F\frac{(J^{3}+\psi^{1}\psi^{2})}{\sqrt{k+2}}\frac{\overline{J}}{\sqrt{k_{g}}}$
where $J^{3}$, $\overline{J}^{3}$ are the $SU(2)$ currents, $J$,
$\overline{J}$ are holomorphic and antiholomorphic ones, and the
right moving current $\overline{J}$ is normalized as $<\overline{J}(1)\overline{J}(0)>=k_{g}/2$.
The corresponding gravitational deformation reads: $V_{gr}={\cal R}\frac{(J^{3}+\psi^{1}\psi^{2})\overline{J}^{3}}{\sqrt{k+2}\sqrt{k}}$.

Let us include these marginal deformations $V_{F}$ and $V_{gr}$
as $O(1,1)$ boost in the lattice of charges of the theory. The effects
will be encoded in the zero-modes of the $SU(2)_{k}$ currents, $J^{3}$,
$\overline{J}^{3}$, i.e. $I$, $\bar{I}$, the zero-modes of the
holo- (antiholo-)morphic currents, $J$, $\overline{J}$, i.e. ${\cal P}$,
$\bar{{\cal P}}$, and the zero-mode of the holomorphic helicity current,
$\psi^{1}\psi^{2}$, which is denoted by ${\cal Q}$. Then the zero-modes
of the algebra are:

\[
L_{0}=\frac{{\cal Q}^{2}}{2}+\frac{I^{2}}{2}+...\,,\;\bar{L}_{0}=\frac{\bar{{\cal P}}^{2}}{k_{g}}+...\]
 which gives rise to the relevant for the $V_{F}$ perturbation part:

\begin{equation}
L_{0}=\frac{({\cal Q}+I)^{2}}{k+2}+\frac{k}{2(k+2)}\left({\cal Q}-\frac{2}{k}I\right)^{2}+...\;.\end{equation}
 The $O(1,1)$ boost mixes the holomorphic zero-mode current $I+{\cal Q}$
with the antiholomorphic $\bar{{\cal P}}$:

\begin{equation}
\begin{array}{c}
L'_{0}=\left(\cosh x\frac{{\cal Q}+I}{\sqrt{k+2}}+\sinh x\frac{\bar{{\cal P}}}{\sqrt{k_{g}}}\right)^{2}+\frac{k}{2(k+2)}\left({\cal Q}-\frac{2}{k}I\right)^{2}+...\\[4pt]
\bar{L}'_{0}=\left(\sinh x\frac{{\cal Q}+I}{\sqrt{k+2}}+\cosh x\frac{\bar{{\cal P}}}{\sqrt{k_{g}}}\right)^{2}+...\;.\end{array}\label{Ap2}\end{equation}
Next we include the $V_{gr}$ deformation. This deformation is symmetric
hence we perform in addition to $V_{F}$ the $O(2)$ transformation
which mixes the antiholomorphic $\bar{I}$, $\bar{{\cal P}}$. After
substituting $F=\sinh(2x)$ we have the following perturbation $\delta L_{0}=L'_{0}-L_{0}$
in $L_{0}$ (and $\bar{L}_{0}$):
\begin{widetext}
\begin{equation}
\delta L_{0}=\left[\frac{{\cal R}\bar{I}}{\sqrt{k}}+\frac{F^{2}}{\sqrt{k_{g}}}\right]\frac{{\cal Q}+I}{\sqrt{k+2}}+\left(\sqrt{1+{\cal R}^{2}+F^{2}}-1\right)\left[\frac{({\cal Q}+I)^{2}}{2(k+2)}+\frac{1}{2({\cal R}^{2}+F^{2})}\left(\frac{{\cal R}\bar{I}}{\sqrt{k}}+\frac{F\bar{{\cal P}}}{\sqrt{k_{g}}}\right)^{2}\right].\end{equation}
\end{widetext}
This perturbation gives the mass spectra $L_{0}=M_{L}^{2}$ ($\bar{L}_{0}=M_{R}^{2}$):

\begin{equation}
\begin{array}{c}
M_{L}^{2}=-\frac{1}{2}+\frac{{\cal Q}^{2}}{2}+\frac{1}{2}\sum_{i=1}^{3}{\cal Q}_{i}^{2}+\frac{(j+1/2)^{2}-({\cal Q}+I)^{2}}{k+2}+E_{0}+\\[4pt]
+\frac{1+\sqrt{1+F^{2}}}{2}\left[\frac{{\cal Q}+I}{\sqrt{k+2}}+\frac{F\bar{{\cal P}}}{\sqrt{k_{g}}\left(1+\sqrt{1+F^{2}}\right)}\right]^{2}+\\[4pt]
+\frac{1+\sqrt{1+{\cal R}^{2}}}{2}\left[\frac{{\cal Q}+I}{\sqrt{k+2}}+\frac{{\cal R}\bar{I}}{\sqrt{k}(1+\sqrt{1+{\cal R}^{2}})}\right]^{2}\end{array}\label{Ap3}\end{equation}
and with the help of the following gravitational backreaction moduli:

\begin{equation}
\lambda=\sqrt{{\cal R}+\sqrt{1+{\cal R}^{2}}},\frac{1}{\lambda}=\sqrt{-{\cal R}+\sqrt{1+{\cal R}^{2}}}\label{eq:moduli-lambda}\end{equation}
 the spectra can be obtained:

\begin{equation}
\begin{array}{c}
M_{L}^{2}=-\frac{1}{2}+\frac{{\cal Q}^{2}}{2}+\frac{1}{2}\sum_{i=1}^{3}{\cal Q}_{i}^{2}+\frac{(j+1/2)^{2}-({\cal Q}+I)^{2}}{k+2}+E_{0}+\\[4pt]
+\frac{1+\sqrt{1+F^{2}}}{2}\left[\frac{{\cal Q}+I}{\sqrt{k+2}}+\frac{F\bar{{\cal P}}}{\sqrt{k_{g}}\left(1+\sqrt{1+F^{2}}\right)}\right]^{2}+\\[4pt]
+\frac{1}{4}\left[\left(\lambda+\frac{1}{\lambda}\right)\frac{{\cal Q}+I}{\sqrt{k+2}}+\left(\lambda+\frac{1}{\lambda}\right)\frac{\bar{I}}{\sqrt{k}}\right]^{2}\;.\end{array}\label{Ap4}\end{equation}
The middle lines in (\ref{Ap3}) and (\ref{Ap4}) express the effect
of the single constant magnetic field while the last lines - the gravitational
backreactions. ${\cal Q}_{i}$, $i=1,2,3$ refer to the helicity operators
corresponding to the internal left fermions in the background $SU(2)_{k}\times\mathbb{R}_{\phi}\times K^{6}$
and $j=0,1,2,...,p$ for $k=2p$.

Similarly, the right moving mass spectra read:

\begin{equation}
\begin{array}{c}
M_{R}^{2}=-1+\frac{P^{2}}{k_{g}}+\frac{(j+1/2)^{2}-({\cal Q}+I)^{2}}{k+2}+\bar{E}_{0}+\\[4pt]
+\frac{1+\sqrt{1+F^{2}}}{2}\left[\frac{{\cal Q}+I}{\sqrt{k+2}}+\frac{F\bar{{\cal P}}}{\sqrt{k_{g}}\left(1+\sqrt{1+F^{2}}\right)}\right]^{2}+\\[4pt]
+\frac{1}{4}\left[\left(\lambda+\frac{1}{\lambda}\right)\frac{{\cal Q}+I}{\sqrt{k+2}}+\left(\lambda+\frac{1}{\lambda}\right)\frac{\bar{I}}{\sqrt{k}}\right]^{2}\;.\end{array}\label{Ap5}\end{equation}
For the completeness let us write the partition function $Z^{W}(\tau,\bar{\tau})$
for the curved 4-spacetime $W=SU(2)_{k=2p}\times\mathbb{R}_{Q}$ in
terms of flat 4-space function $Z_{0}$ (still for $k=2p$):

\begin{equation}
Z^{W}(\tau,\bar{\tau})={\rm Im}\tau^{3/2}|\eta(\tau)|^{6}\frac{\sum_{\gamma,\delta=0}^{1}Z_{so(3)}\left[\begin{array}{c}
\gamma\\
\delta\end{array}\right]}{2V}Z_{0}(\tau,\bar{\tau})\label{eq:Z-curved}\end{equation}
 and $Z_{so(3)}\left[\begin{array}{c}
\gamma\\
\delta\end{array}\right]=e^{-i\pi\gamma\delta k/2}\sum_{l=0}^{k}e^{i\pi\delta l}\chi_{l}\bar{\chi_{(1-2\gamma)l+\gamma k}}$ where $\chi_{l}(\tau)$ are the $SU(2)_{k}$ characters:

\begin{equation}
\chi_{l}(\tau)=\frac{\vartheta_{l+1,k+2}(\tau,\upsilon)-\vartheta_{-l-1,k+2}(\tau,\upsilon)}{\vartheta_{1,2}(\tau,\upsilon)-\vartheta_{-1,2}(\tau,\upsilon)}|_{\upsilon=0}\end{equation}
 with the level $k$ $\vartheta$-functions.

Together with the mass spectra (\ref{Ap4}), (\ref{Ap5}) we have
the complete and exact string spectra in the presence of the constant
magnetic field and curvature ${\cal R}$. Then, based on the mass
spectra and (\ref{eq:Spectrum G+H}) one derives the dictionary rules
as in (\ref{Dic2}) \cite{KK95}, where $\lambda^{2}\approx1+{\cal R}+{\cal O}(F^{2},{\cal R}^{2})$
which agrees with (\ref{eq:moduli-lambda}) up to ${\cal O}(F^{2})$.

\end{document}